# Efficient heralding of O-band passively spatial-multiplexed photons for noise-tolerant quantum key distribution

**Mao Tong Liu[1] and Han Chuen Lim[1,2,*]**

[1]*School of Physical and Mathematical Sciences, Nanyang Technological University, 21, Nanyang Link, 637371, Singapore*
[2]*Emerging Systems Division, DSO National Laboratories, 20, Science Park Drive, 118230, Singapore*
[*]*hanchuen@dso.org.sg*

**Abstract:** When implementing O-band quantum key distribution on optical fiber transmission lines carrying C-band data traffic, noise photons that arise from spontaneous Raman scattering or insufficient filtering of the classical data channels could cause the quantum bit-error rate to exceed the security threshold. In this case, a photon heralding scheme may be used to reject the uncorrelated noise photons in order to restore the quantum bit-error rate to a low level. However, the secure key rate would suffer unless one uses a heralded photon source with sufficiently high heralding rate and heralding efficiency. In this work we demonstrate a heralded photon source that has a heralding efficiency that is as high as 74.5%. One disadvantage of a typical heralded photon source is that the long deadtime of the heralding detector results in a significant drop in the heralding rate. To counter this problem, we propose a passively spatial-multiplexed configuration at the heralding arm. Using two heralding detectors in this configuration, we obtain an increase in the heralding rate by 37% and a corresponding increase in the heralded photon detection rate by 16%. We transmitted the O-band photons over 10 km of noisy optical fiber to observe the relation between quantum bit-error rate and noise-degraded second-order correlation function of the transmitted photons. The effects of afterpulsing when we shorten the deadtime of the heralding detectors are also observed and discussed.

## 1. Introduction

A well-implemented quantum key distribution (QKD) system in combination with one-time-pad encryption can offer communication security that is free from computational assumptions [1, 2]. Despite this, QKD has not found widespread deployment in commercial fiber-optic networks yet. Apart from limited key rates and transmission distances, another perceived shortcoming of QKD is that it operates only on virtually-noise-free dark fibers [3–5], which are costly. It is therefore of interest to extend the operation of QKD to optical fiber lines carrying other types of optical signals [6–12].

Recent experiments have shown significant progress in this direction. Patel *et al.* have employed tight temporal gating of GHz-clocked photon detectors to achieve wavelength-multiplexing of QKD and classical data channels within the low-loss C-band (1530–1565 nm) for transmission distances of up to 70 km [7, 8]. For shorter transmission distances, one might consider transmitting QKD photons in the O-band (1260–1360 nm) where the noise condition is less severe [9–11]. A photon heralding scheme for noise-tolerant O-band QKD has been suggested [12]. In this scheme, the photons are produced from a heralded photon source (HPS) or a heralded *single*-photon source when multi-photon components can be neglected.

An HPS is a photon-pair source in which one photon of each photon-pair is detected at the source to trigger a heralding signal that could be used to *herald* the presence of the other photon of the same pair [13–19]. As the detected photon plays a heralding role, it is called the *heralding photon*. The photon that is being heralded is called the *heralded photon*. In the suggested scheme, heralding signals are transmitted together with the heralded photons and used to trigger the photon detectors at the receiver such that they are turned on only during time-slots in which a heralded photon is expected to arrive.

For convenience, we introduce a quantity called *photon signal-to-noise ratio* (PSNR), which is defined as the ratio between probability of detecting a useful photon and probability of detecting a noise photon at the receiver. When applied to QKD, PSNR is related to quantum bit-error rate (QBER) by

$$QBER \approx \frac{1}{2(1+PSNR)}, \qquad (1)$$

where it is assumed that the noise is basis-independent and distributed equally over all the photon detectors, and that there is no more than one detected noise photon per time-slot [12]. We have shown in [12] that the use of HPS instead of a weak coherent source (WCS) for QKD could lead to an increased PSNR and hence lower QBER according to (1).

The realization of noise-tolerant O-band QKD based on photon heralding could enable non-disruptive introduction of QKD into existing optical fiber networks without any wavelength or power restrictions imposed on current Internet data services. However, the key rate might suffer as the photon transmission rate in a photon heralding scheme is directly proportional to the product of *heralding probability* and *heralding efficiency* of the HPS, and both are < 1 for a typical HPS. We define heralding probability as the probability of obtaining a heralding signal per time-slot, and heralding efficiency as the probability of finding a non-empty time-slot at the transmitter output when there is a heralding signal. It is thus important to improve both heralding probability and heralding efficiency in an HPS in order to achieve a high key rate.

There are already reports on 1310 nm photon-pair sources based on different materials and configurations [20–24]. We choose symmetric spontaneous parametric down-conversion (SPDC) in a periodically-poled lithium niobate (PPLN) waveguide [25–34] as our source of O-band photon-pairs because of its high SPDC efficiency and low output coupling loss. The heralding efficiency of our HPS is measured to be 74.5%. This is much higher than the value of 22.4% obtained using our previous source of O-band photon-pairs based on a silicon wire waveguide. For the choice of mean photon-pair number $\mu = 0.12$, using our HPS for photon transmission improves PSNR by a factor of $\chi = 6.95$ as compared to using WCS. This value is calculated from $\chi = \alpha_s (1+\mu)/\mu$, where $\alpha_s$ is approximately the heralding efficiency [12].

The heralding probability depends on the optical loss of the heralding arm and the quantum efficiency of the *heralding detector* that detects the heralding photons. For photon detectors based on InGaAs avalanche photodiodes (APDs), it is common to set a deadtime of 10 μs in order to suppress afterpulsing. However, we find that this significantly reduces heralding rate of the HPS as heralding photons that arrive within the deadtime periods are not detected.

In this work, we implement a passive spatial-multiplexing (PSM) scheme to counter this problem. Differing from active-multiplexing schemes that optically switch the output of many parametric down-conversion sources to increase single-photon probability [35–42], the purpose of having PSM is simply to reduce the number of undetected heralding photons due to a long detector deadtime [43, 44]. In our experiment, we use a 1×2 optical fiber coupler and two heralding detectors to construct a 1×2 PSM configuration to demonstrate the PSM concept.

The main advantage of PSM is in its simplicity and low loss as it uses an optical splitter. The performance of the PSM scheme should approach that of a photon-number resolving detector in the limit of large number of heralding detectors and negligible optical loss of the heralding arm [45]. PSM offers an opportunity for partial identification of the time-slots that contain multi-pairs. This is because whenever two heralding detectors register detection simultaneously, it is likely caused by multiple-pair emission. In this case, the heralding signal can be removed to reduce multi-photon probability at the heralded side. It should be pointed out here that the active spatial-multiplexing scheme of Migdall *et al.* [35] maintains a high single-photon fraction even when heralding arm has high optical loss because each photon-pair source in the array satisfies $\mu \ll 1$.

In our experiment, we have used an exclusive-or (XOR) circuit at the output of the two heralding detectors to remove the heralding signal whenever there is coincidence detection at the two heralding detectors. However, due to optical loss at the heralding arm and non-unity quantum efficiency of the heralding detectors, multi-photon components cannot be eliminated completely. For a QKD system, it would be necessary to estimate the multi-photon probability and take this into account when performing privacy amplification.

The rest of this paper is organized as follows. Section 2 describes the characterization of our O-band photon-pair source. Section 3 explains the concept of PSM and presents key experimental results showing improvements to heralding rate and photon detection rate at the receiver. In Section 4, we transmit the O-band photons over 10 km of noisy optical fiber to observe the relation between QBER and noise-degraded second-order correlation function, $g^{(2)}(0)$, of the transmitted photons. Section 5 investigates the effect of setting a shorter heralding detector deadtime. Section 6 is a discussion on the benefits of using photon heralding and PSM. Section 7 concludes this work. A theoretical treatment of two-detector PSM scheme is provided in Appendix A.

## 2. Photon-pair source

We produce O-band photon-pairs over a broad bandwidth using two PPLN waveguides (HC Photonics) in a cascade configuration [30–32], as shown in Fig. 1.

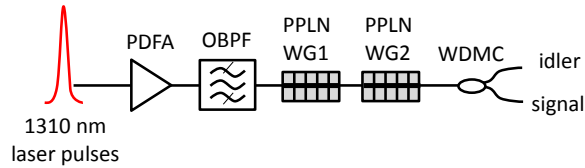

Fig. 1. Schematic of our photon-pair source based on two PPLN waveguides. The photon pairs are traditionally called *signal* and *idler*. For measurement of SPDC spectral half-width, we attach an arrayed waveguide grating (AWG) to the signal arm to select the photon wavelength. For coincidence-to-accidental ratio (CAR) measurement, we use two wavelength-tunable fiber Bragg grating filters to select the signal and idler photons. OBPF: optical band-pass filter; PDFA: praseodymium-doped fiber amplifier; PPLNWG: periodically-poled lithium niobate waveguide; WDMC: wavelength-division-multiplexing coupler.

The first PPLN waveguide is 15-mm-long and it is pumped by 1310 nm laser pulses to produce 655 nm pulses via second harmonic generation (SHG). The pump pulses have pulse-widths of 100 ps and a pulse repetition rate of 48.7 MHz. They are optically amplified by a praseodymium-doped fiber amplifier (PDFA, FiberLabs, AMP-FL8611-OB-20). An optical band-pass filter (OBPF, Yenista Optics, XTM-50) serves to filter off amplified spontaneous emission (ASE) noise coming from the PDFA. The 655 nm pulses produced by the first PPLN waveguide are directed to the second PPLN waveguide, which is 5-mm-long, to produce O-band photon-pairs over a broad bandwidth via symmetric SPDC. We characterize our source by measuring the half-spectrum and coincidence-to-accidental ratio (CAR) of the output photon-pairs.

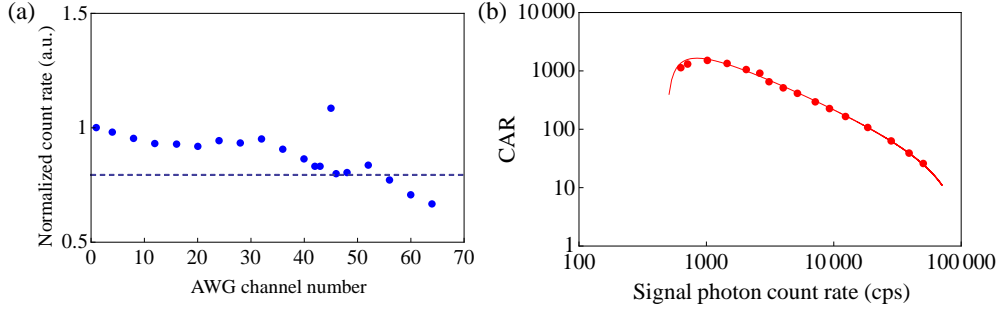

Fig. 2. Photon-pair source characterization results. (a) Measured photon-pair generation half-spectrum which combines SPDC source spectrum and AWG channel transmittance. The effect of detector deadtime has been corrected using Eq. (3) in Section 3. For normalization, all count rates are divided by Channel 1's count rate. The higher count rate at Channel 45 is due to leakage of 655 nm photons through a free-spectral range (FSR) mode of the AWG. The broken line marks the –1 dB level which shows a measured half-spectral-width of 15.5 nm. (b) Measured CAR values versus count rates at signal wavelength of 1301.75 nm (corresponding to Channel 25 of our AWG), as pump power is changed. The solid curve is a theoretical fit obtained using experimental parameters of heralding arm optical loss of –13.0 dB, detector deadtime of 10 μs, and detector dark count rates of $5.5\times10^{-6}$ (idler detector) and $1.0\times10^{-5}$ (signal detector).

Figure 2(a) shows the normalized signal-band photon count rates measured with a custom-made 64-channel O-band arrayed waveguide grating (AWG). The AWG has a channel spacing of 50 GHz and it covers a wavelength range from 1308.57 nm (Channel 1) to 1290.82 nm (Channel 64). We use a gated single-photon counter module (SPCM, IDQ id210) for the measurement. The 1-dB spectral half-width is measured to be 15.5 nm, which suggests that the source is suitable for generating O-band photon-pairs over a broad bandwidth.

The quality of the photon-pairs is evaluated by measuring the CAR under different pump powers. In this experiment, we detect the signal and idler photons using two gated SPCMs (IDQ, id210). Two tunable fiber Bragg grating (FBG) filters (AOS, TFBG131x-300) are used to select the idler and the signal channels separately. The SPCM outputs are sent to an electronic two-fold-coincidence module (Kaizuworks, KN1250) for coincidence-counting. Figure 2(b) shows measurement result obtained for signal wavelength of 1301.75 nm (corresponding to Channel 25 of the AWG). At a signal photon count rate of about 1 kcps, we observe the highest CAR value of 1518. The average pump power incident on the first PPLN waveguide is –6.5 dBm. This shows that our source produces very high quality photon-pairs. All the following experiments are carried out using Channel 25.

## 3. Passively spatial-multiplexed heralding

A typical HPS uses just one heralding detector. Detector deadtime suppresses afterpulsing of the photon detector but it also reduces the probability of detecting heralding photons. In the absence of detector deadtime, the *triggering rate*, which is the number of triggers produced by the heralding detector per second, can be expressed as

$$N_t = [1-(1-d_t)\exp(-\beta\mu)]F . \qquad (2)$$

where $\beta$ is the photon collection and detection efficiency that takes into account the optical transmittance of the heralding arm and quantum efficiency of the heralding detector. $\mu$ is the mean photon-pair number, $F$ is the detector gating rate, and $d_t$ is the detector dark count probability per gate. The heralding detector has a deadtime of $\tau_{hr}$, and therefore the rate of detection events is reduced to

$$N_d = \frac{N_t}{1+N_t\tau_{hr}} . \qquad (3)$$

The idea of PSM is to use a 1×m optical splitter or fiber coupler to divide the heralding arm into m spatially-multiplexed output ports and to use m photon detectors each having a deadtime of $\tau_{hr}$ to detect the heralding photons in parallel. Assuming that all the m photon detectors are identical and that the m spatially-multiplexed arms have the same optical loss, the overall heralding rate $N_{d,m}$ can be expressed as

$$N_{d,m} = \frac{\left[1-(1-d_t)\exp\left(-\frac{\beta\mu}{m}\right)\right]Fm}{1+\left[1-(1-d_t)\exp\left(-\frac{\beta\mu}{m}\right)\right]F\tau_{hr}}. \quad (4)$$

To illustrate the benefits of PSM, let us consider $N_t$ = 300 kcps, $\tau_{hr}$ = 10 μs, and assuming $d_t$ = 0. With just one heralding detector, we obtain a heralding rate of $N_d$ = 75 kcps, which is only 25% of the heralding rate in the absence of deadtime. If we use m = 2 in PSM configuration, the heralding rate can be improved to $N_{d,2}$ = 120 kcps. This is 40% of $N_t$ and 60% higher than $N_d$. Using more heralding detectors will increase the percentage but implementation cost becomes higher.

PSM also provides an opportunity for partial identification of the time-slots that contain multi-pairs [45]. This is because whenever two or more of the m detectors register detection simultaneously, it is likely caused by multi-pair-emission. In this case, the heralding signal can be removed to reduce multi-pair probability.

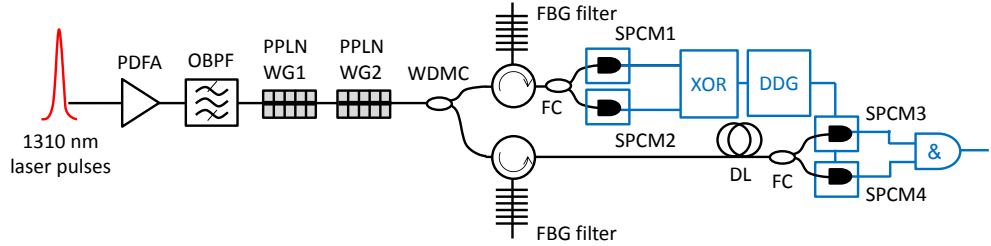

Fig. 3. Schematic of the experiment. SPCMs 1 and 2 are passively-spatial-multiplexed by using a 1×2 optical fiber coupler. SPCMs 3 and 4 perform a Hanbury-Brown–Twiss type measurement. DDG: digital delay generator; DL: delay line; FBG: fiber Bragg grating; FC: fiber coupler; SPCM: single-photon counter module; XOR: exclusive-or circuit.

The second-order correlation function, $g^{(2)}(0)$, can be used to characterize the quality of the heralded photons. Ideally, $g^{(2)}(0)$ = 0 for a perfect HPS that outputs only single-photons. In general, due to multi-photon components in the SPDC process and presence of noise photons, $g^{(2)}(0)$ must be > 0. We would like to observe how PSM improves both heralding rate and heralded photon detection rate for the same measured $g^{(2)}(0)$ value. In order to do this, we need to perform a Hanbury-Brown–Twiss (HBT) type three-fold-coincidence experiment. Figure 3 shows the experimental schematic. For this experiment, we have used two tunable FBG filters (AOS, TFBG131x-300) having full-width half-maximum (FWHM) pass-band of 0.3 ± 0.05 nm to select correlated photon-pairs. Pump photons are largely removed when coupling 655 nm SHG light into the second PPLN waveguide. Residual pump remains at the output port of the second PPLN waveguide. Its rejection is achieved by the two reflection-type FBG filters that we use at the receiver to select the correlated photon-pairs. The low out-coupling loss of the PPLN waveguide and the removal of pump rejection FBGs have resulted in a high heralding efficiency of this HPS.

We perform two sets of measurement. For the first set, we remove both the 1×2 coupler and SPCM 2 from the experimental setup and use only one heralding detector, SPCM 1 (IDQ, id210), to trigger two gated SPCMs 3 and 4 (IDQ, id201) to measure $g^{(2)}(0)$. We use id210 at heralding arm because its maximum trigger frequency is 100 MHz. We trigger it at 48.7 MHz and use it to detect heralding photons. On the other hand, the maximum trigger frequency for id201 is 8 MHz, and so we use it at the receiver side where triggering frequency is < 8 MHz. All SPCM deadtimes are set to 10 µs to minimize afterpulsing. We use the method described in Appendix B of [12] to find the optical loss of the heralding arm. When $g^{(2)}(0) = 0.2$, the mean photon-pair number of the HPS is found to be $\mu = 0.12$. Using measured values of the optical losses of FBG filters, delay lines, and calibrated SPCM quantum efficiencies, we obtain a heralding efficiency value of 74.5% for our HPS. Using this HPS for photon transmission at $\mu = 0.12$, we can expect a PSNR that is 6.95 times higher compared to using WCS. This is much higher than the value of 2.26 obtained in [12].

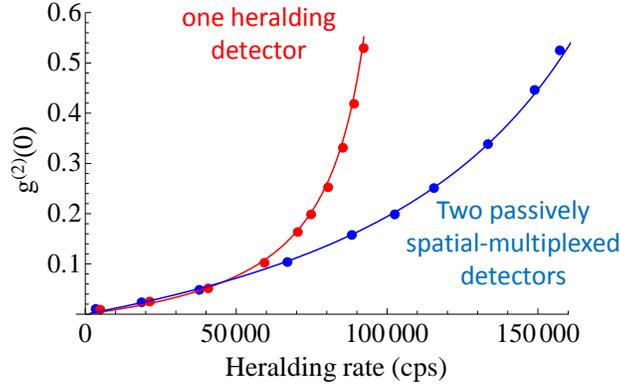

Fig. 4. Measured $g^{(2)}(0)$ values versus heralding rates for the cases of one heralding detector (in red) and two passively-spatial-multiplexed heralding detectors (in blue). The error bars are all smaller than the size of the markers. The two solid curves are theoretical fitting curves.

For the second set of measurement, we use two heralding detectors, SPCMs 1 and 2 (IDQ, id210) in PSM configuration, as shown in Fig. 3. We further include a complex programmable logic device (CPLD, Lattice Semiconductors, Brevia2 development kit) to implement an XOR circuit. For this set of measurement, the XOR output is used to trigger SPCMs 3 and 4. We apply a deadtime of 120 ns to the XOR circuit because our digital delay generator (DDG, Highland Technology, P400) does not work with pulse repetition rates of >10 MHz.

The $g^{(2)}(0)$ values obtained under different pump powers for both sets of measurements are plotted against measured heralding rate in Fig. 4 for comparison. As expected, a higher heralding rate is obtained with two heralding detectors in PSM configuration. When average pump power is 7.5 dBm, we can obtain $g^{(2)}(0)$ of 0.2. At this time, using just one heralding detector, we detected $r = 74,843$ photons per second. This is only about 25% of the total number of heralding photons that would have been detected if the photon detector had no deadtime. Using two heralding detectors in PSM configuration, the heralding rate is increased by 37%. This is lower than the 60% increase predicted using Eq. (4) mainly because of the insertion loss of the 1×2 optical fiber coupler, and a slightly lower quantum efficiency of the second heralding detector.

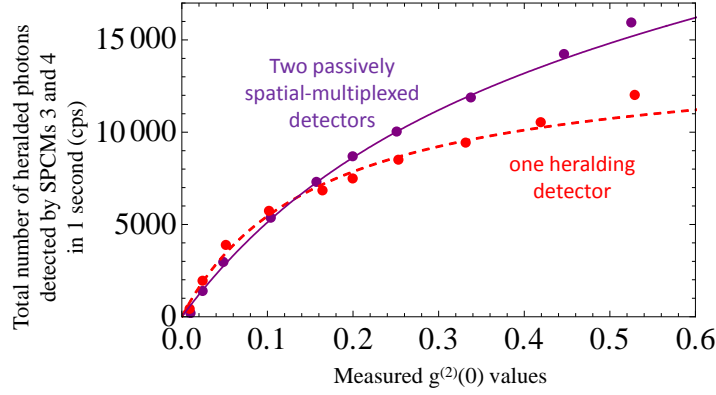

Fig. 5. The total number of heralded photons detected by SPCMs 3 and 4 within 1 second versus measured $g^{(2)}(0)$ values. Filled squares are for the case of a single heralding detector, SPCM 1. Empty circles are for the case of using two passively-spatial-multiplexed heralding detectors, SPCMs 1 and 2. The error bars are all smaller than the size of the markers. The solid and broken curves are theoretical fitting curves.

The total detection count rate of SPCMs 3 and 4 is plotted against $g^{(2)}(0)$ values in Fig. 5. At $g^{(2)}(0) = 0.2$, the two SPCMs register a total of 8,707 detection events per second. This is a 16% increase from the 7,502 detections per second when there is only one heralding detector. The photon detection rate at the receiver does not increase by the same percentage as the heralding rate, because the deadtime of the receiver SPCMs blocks some of the heralded photons. This is in agreement with theory, which is explained in Appendix A.

## 4. Transmission experiment

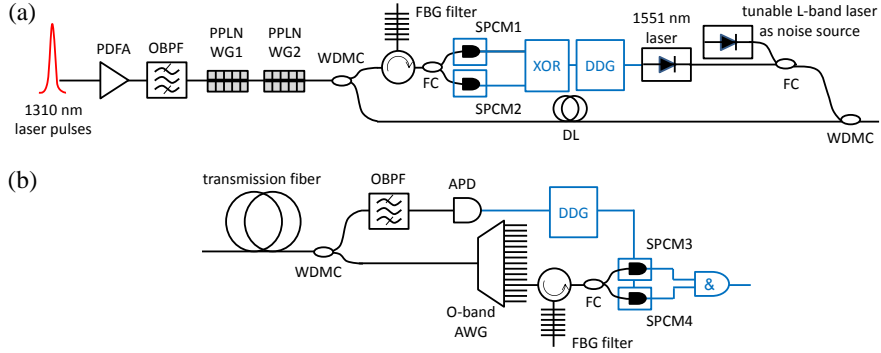

Fig. 6. (a) Schematic of the transmitter. It consists of a photon-pair source with two passively-spatial-multiplexed heralding detectors and a 1551 nm laser to produce heralding signal pulses. A tunable L-band laser is used as noise source. (b) Schematic of the receiver, which implements a Hanbury-Brown–Twiss type measurement. APD: avalanche photodiode; AWG: arrayed waveguide grating.

Using the PSM scheme, we have transmitted O-band photons over 10 km of noise-corrupted optical fiber. Figure 6(a) shows a schematic of the transmitter, while Fig. 6(b) is a schematic of the receiver. We use a wavelength-tunable L-band laser as a noise source. The wavelength of the L-band laser coincides with a free-spectral range (FSR) mode of the AWG and so some of the L-band photons leak through the AWG and are detected as noise photons at the receiver. This emulates the real-life situation where insufficient optical filtering of classical channels seriously affects the PSNR of the quantum channel. An extra FBG filter has been employed after the AWG to further filter off the L-band noise photons. Inserting more optical filters would lead to greater loss of the QKD photons.

The XOR output triggers a semiconductor laser (IDQ, id300) that produces heralding signal pulses at 1551 nm. The pulses are transmitted together with the O-band photons. At the receiver, the heralding signal pulses are separated from the O-band photons before being filtered and detected by an APD to produce electrical pulses that trigger the two receiver SPCMs. Through adjusting the output power of the L-band laser, we can vary the amount of noise introduced into the optical fiber transmission line and measure both PSNR and $g^{(2)}(0)$ at the receiver. We measure the receiver count rates when both signal source and noise source are turned on to obtain $S' + N$, and when only signal source is turned on (noise source is turned off) to obtain $S$, where $S'$ and $S$ are signal count rates and $N$ is noise count rate. As detector deadtime is set to 10 μs, we can take afterpulsing to be negligible. Noise-photon-detection-induced detector deadtime may make $S'$ slightly smaller than $S$, but we just neglect this effect to obtain PSNR simply from $S / (S' + N - S)$ as it is a good approximation to $S / N$. From PSNR, we can calculate the expected QBER using Eq. (1).

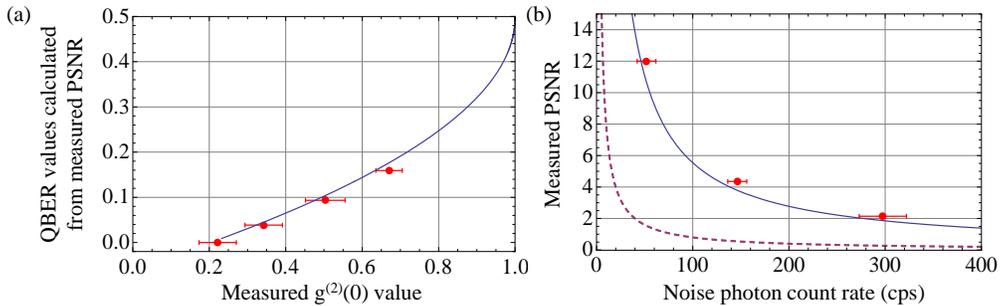

Fig. 7. (a) Observed relation between QBER and noise-degraded $g^{(2)}(0)$ of the heralded photons. QBER values are calculated from measured PSNR after transmission of the O-band heralded photons over 10 km of noisy optical fiber. The different $g^{(2)}(0)$ values were obtained by varying the amount of noise launched into the transmission fiber. The solid curve is theoretical estimation. Error bars for QBER are smaller than the size of the markers. (b) Measured PSNR vs noise photon count rates at receiver. The solid curve is theoretical estimation for our experiment. The broken curve is what theory predicts for WCS. The parameters used are $\alpha_s = 0.745$ and $\mu = 0.12$, which gives $\chi = 6.95$.

The results are plotted in Fig. 7(a), which clearly shows the relation between QBER and noise-degraded $g^{(2)}(0)$. In the case where the noise source is switched off, $g^{(2)}(0)$ is approximately 0.2 and we assume QBER = 0 since we have not considered other system imperfections such as basis misalignments. We have also neglected detector dark counts as its contribution is < 1% of the photon detection rates. We see that QBER < 0.10 for $g^{(2)}(0) < 0.5$. The increase of $g^{(2)}(0)$ value from approximately 0.2 to 0.5 and higher is due to addition of noise photons, and so this increase does not imply that the system leaks more information to eavesdroppers, as would have been the case if the SPDC source emits photons with a high $g^{(2)}(0)$ value such as 0.5. It is therefore appropriate for us to use the security threshold for QBER. It is known that for QBER values exceeding 0.11, no secure key can be generated.

Figure 7(b) shows a plot of measured PSNR versus noise photon count rates at the receiver. The broken curve is theoretically predicted PSNR curve for WCS with mean photon-number of 0.12. When PSNR is < 4, the corresponding QBER value exceeds 0.10. This result demonstrates the benefit of using photon heralding over WCS under noisy conditions.

## 5. Effect of shortening heralding detector deadtime

Next, we investigate the effect of shortening the deadtime of heralding detectors. A shorter deadtime increases the heralding rate because fewer heralding photons will be blocked by detector deadtime but at the same time, increased afterpulsing may lead to some adverse effects. For heralding detector deadtime settings of 10, 5 and 1 µs, we measure the heralding rates and the photon detection rates at the receiver when a noise source is switched on and off. The noise source is the same L-band laser that we used in the transmission experiment. For simplicity and avoidance of complication due to other transmission issues, we do not use a transmission fiber for this experiment. For the purpose of comparison, the measurements are performed with and without an attenuation of –3.36 dB applied to the heralded photons. The deadtimes of the photon detectors at the receiver are set to 10 µs for all measurements. Figure 8 shows the experimental schematic.

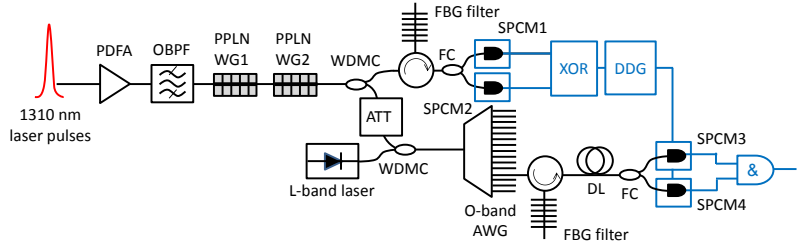

Fig. 8. Schematic of the experiment on shortening heralding detector deadtime. The attenuation value used for the attenuator (ATT) is –3.36 dB.

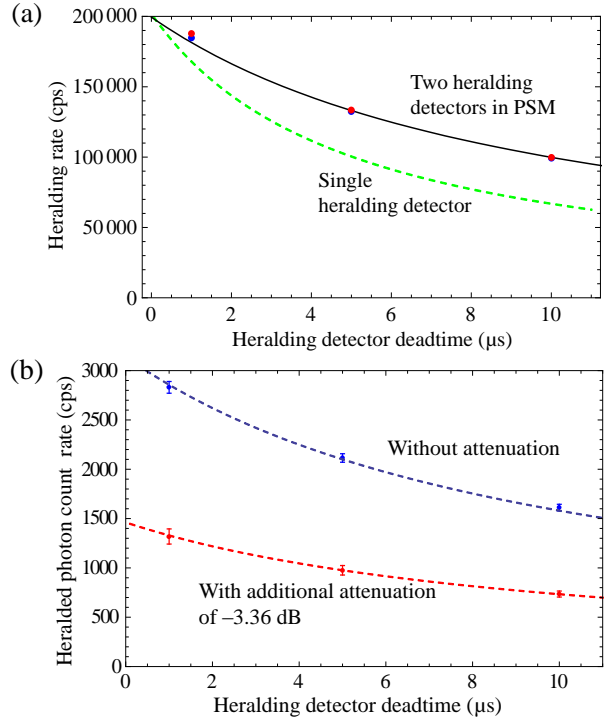

Fig. 9. (a) Heralding rate versus heralding detector deadtime. Black solid curve is theoretical result obtained for two heralding detectors in PSM configuration. The broken curve (in green) is theoretical result for the case of single heralding detector. The theoretical curves are obtained using Eq. (4). The red and blue dots represent measurements on heralding rates for the two cases of with and without –3.36 dB attenuation. The error bars are all smaller than the size of the dots. (b) Heralded photon count rates at the receiver for the two cases of with and without –3.36 dB attenuation. Dashed curves are theoretical curves obtained from Eq. (16) as given in Appendix A and fitted to the experimental data.

Figure 9(a) shows that using two heralding detectors in PSM configuration instead of one leads to a higher heralding rate. When the heralding detectors' deadtimes are set to 1 μs, we can obtain a heralding rate that is > 90% of that in the absence of deadtime, but we should also expect more afterpulsing. Figure 9(b) shows that in the case of two-detector PSM, higher detection count rates at the receiver can be obtained for shorter heralding detector deadtimes.

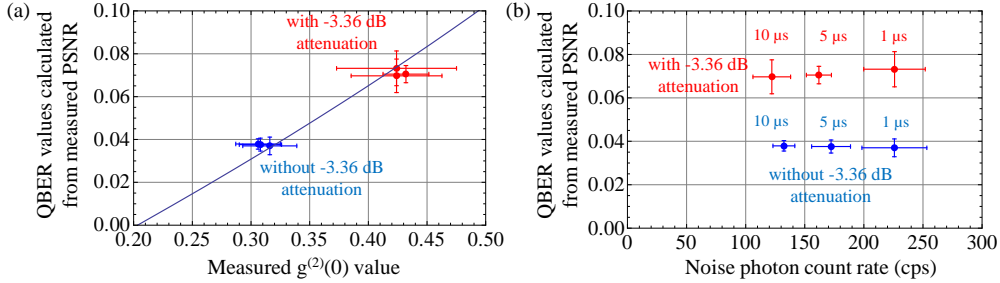

Fig. 10(a). QBER calculated from measured PSNR versus measured $g^{(2)}(0)$ values for different heralding detector deadtime settings, and for both cases of without (blue) and with (red) -3.36 dB attenuation. The solid curve is the same as the theoretical curve shown in Fig. 7(a) in the previous section. (b) QBER calculated from measured PSNR versus measured noise photon count rates at receiver. For both cases of without (blue) and with (red) −3.36 dB attenuation, the QBER values and noise photon count rates for three different heralding detector deadtimes of 1, 5, and 10 μs are shown.

Figure 10(a) compares the QBER and $g^{(2)}(0)$ values for detector deadtime settings of 1, 5, and 10 μs with and without the additional −3.36 dB attenuation. With additional attenuation, both QBER and $g^{(2)}(0)$ are degraded. This is due to reduced number of signal photons while noise source is kept unchanged. The result also shows that changing heralding detector deadtime does not have significant effect on QBER and $g^{(2)}(0)$ values, despite that more afterpulsing of the heralding detectors can be expected. Figure 10(b) shows that decreasing heralding detector deadtime increases noise photon count rate but this does not affect QBER significantly as signal detection rate is increased as well. Note that the standard deviations of all the measurements are dominated by large fluctuations of the noise photon count rates.

A possible explanation for this observation goes as follows. In the presence of afterpulsing at heralding detector, there are additional afterpulse-heralded signal photons and afterpulse-heralded noise photons being detected at the receiver. The PSNR in the presence of afterpulsing can be expressed as $(S + S_{AP}) / (N + N_{AP})$, where $S$ and $N$ are the signal and noise photon detection rates in the absence of afterpulsing, $S_{AP}$ and $N_{AP}$ are the afterpulse-heralded signal and noise photon detection rates.

Since afterpulses trigger signal photon or noise photon or nothing randomly, the ratio $S_{AP}/N_{AP}$ is likely to be approximately equal to the proportion of signal photons to noise photons present in the channel, which is given by $S / (N \chi)$. As $\chi = 6.95$ for our HPS and assuming $N_{AP} \ll N \chi$, PSNR can be approximated to be $S / (N + N_{AP})$. We suspect that $N_{AP}$ is < 10% of $N$ even for heralding detector deadtime of 1 μs in our experiment, and therefore PSNR is not affected significantly by the afterpulsing at the heralding detector. Under this condition, in practical noise-tolerant QKD using heralded photons, setting a shorter heralding detector deadtime might be an effective way to provide a higher heralding rate while maintaining QBER. This should be verified in future work. However, it should be understood that this method would be ineffective for entanglement-based QKD since afterpulses are not correlated to the heralded photons in any way.

## 6. Discussion

To achieve noise-tolerance, we make use of the temporal correlation between the heralding and the heralded photons. The advantage of physically transmitting the heralding signals over optical transmission line to trigger the photon detectors at the receiver is that in the presence of noise, detector deadtimes due to detection of noise photons can be avoided by this method. We can calculate the improvements brought about by this method. In the presence of noise photon detection rate $N_{noise}$, the QKD photon detection rate at the receiver can be expressed as

$$N_{QKD,noisy} = \frac{N_{QKD}}{1+(N_{QKD}+N_{noise})\tau_{hd}}, \quad (5)$$

where $N_{QKD}$ is the QKD photon detection rate in the absence of noise. Compared with the ideal case of noiseless transmission, the fraction of QKD photons detected is given as

$$f = \frac{1+N_{QKD}\tau_{hd}}{1+N_{QKD}\left(1+\frac{1}{PSNR}\right)\tau_{hd}}. \quad (6)$$

It has been shown in [12] that the use of photon heralding improves PSNR by a factor of $\chi$. If the photon detectors at QKD receiver are triggered by heralding signals, the number of noise photons being detected at the QKD receiver will be reduced by $\chi$ and the fraction is improved to

$$f_{improved} = \frac{1+N_{QKD}\tau_{hd}}{1+N_{QKD}\left(1+\frac{1}{\chi\,PSNR}\right)\tau_{hd}}. \quad (7)$$

We have shown in the previous section that setting a shorter deadtime for the heralding detectors does not affect QBER and $g^{(2)}(0)$ significantly in the operation regime of interest. It is therefore tempting to set a shorter deadtime for the photon detectors at the QKD receiver as well. Although a proportion of the afterpulses at the QKD receiver may not be detected due to absence of heralding signals, those that are heralded by chance may still lead to a degradation of the QBER, especially when the heralding rate is high. This degradation of QBER due to afterpulsing at the QKD receiver will not be reflected in our type of experiment because afterpulsing at the receiver is linked to the detection of QKD photons, and we would not be able to differentiate afterpulses from QKD photons. Thus, the PSNR obtained in our type of experiment cannot be used to estimate the QBER correctly for this particular case.

It may be advantageous to use asymmetric SPDC to produce photon-pairs consisting of one visible photon and one O-band photon per pair, such that better quality silicon photon detectors can be used to detect the visible photons for heralding [46, 47]. The use of PSM would be beneficial even for this case. The deadtime of a silicon photon detector is typically about 40-50 ns. This means that a silicon detector cannot detect photons at higher than a rate of 20-25 Mcps. Assuming that future asymmetric SPDC sources can produce visible/1310 nm photon-pairs at high rates of >100 MHz, using two or more silicon photon detectors in PSM configuration for heralding will still improve the heralding rate. Our claim that PSM is advantageous over single heralding detector thus remains unchanged even for asymmetric SPDC sources.

Lastly, it is worth mentioning that the broadband photon-pair output from our HPS may be exploited for large-capacity wavelength-division-multiplexed noise-tolerant QKD [48]. More work will be needed to study the design and feasibility of such a scheme.

## 7. Conclusion

We have demonstrated efficient heralding of O-band photons using an HPS with a heralding efficiency that is as high as 74.5%. Compared with using WCS, using our HPS offers a higher PSNR by 6.95 times, which suggests that it could be used to make QKD more noise-tolerant. The heralding rate is, however, limited by deadtime of the heralding detector deadtime, in addition to optical loss of the heralding arm and quantum efficiency of the heralding detector.

In this work, we have shown the effectiveness of using two-detector PSM configuration to increase the heralding rate and corresponding detection rate of heralded photons. The O-band heralded photons have been transmitted over 10 km of noise-corrupted optical fiber for us to observe the relation between QBER and noise-degraded $g^{(2)}(0)$ of the transmitted photons. We have also investigated the effect of shortening heralding detector's deadtime to improve both the heralding rate and detection rate of heralded photons at the receiver. We observed no significant degradation to PSNR and QBER caused by more afterpulsing at the heralding detectors. For future work, our HPS with high heralding efficiency together with the PSM scheme can be integrated into a photon heralding QKD system to demonstrate noise-tolerant QKD in the O-band. This shall be one step towards non-disruptive introduction of QKD into noisy optical fiber networks that carry Internet data traffic.

**Appendix A: Model for calculating photon detection rate at receiver**

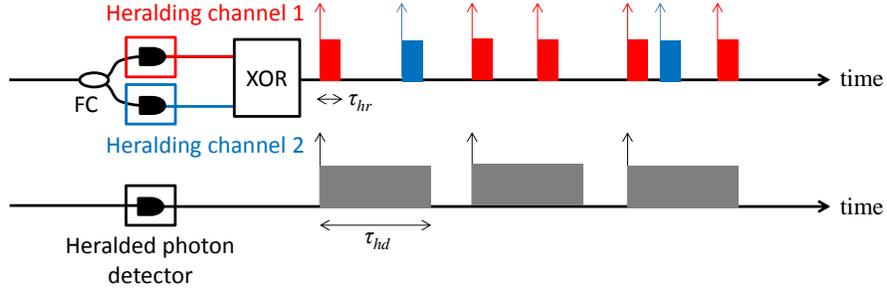

Fig. 11. Model for calculating photon detection rate at the receiver. Vertical arrows indicate photon detection timings. Shaded rectangles indicate deadtime durations. Some photon detection timings at the heralded photon detector fall within detector deadtime durations.

In this appendix, we derive the formulas for the heralding rate and the heralded photon detection rate of our two-detector passively spatial-multiplexed photon heralding scheme. Figure 11 shows the model that we use for the derivation. Neglecting photon detector dark counts for simplicity, the expected photon detection rates at heralding channel $i = 1, 2$, in the absence of deadtime, can be expressed as

$$N_{ti} = \left(1 - e^{-\beta_i \mu}\right) F, \qquad (8)$$

where $\mu$ is the mean photon-pair number, $\beta_i$ is photon collection and detection efficiency of heralding channel $i$, and $F$ is the system clock rate. Taking into account heralding detector's deadtime of $\tau_{hr}$, we can write the photon detection rates as

$$N_{di} = \frac{N_{ti}}{1 + N_{ti} \tau_{hr}}, \qquad (9)$$

The heralding rate is therefore

$$N_{\text{trigger}} = N_{d1} + N_{d2} - 2 N_{\text{XOR}}, \qquad (10)$$

where

$$N_{\text{XOR}} = \frac{N_{d1} N_{d2}}{F} \qquad (11)$$

is the number of coincidence detections in unit time.

To calculate the photon detection rate at the receiver, we have to take into consideration the effect of receiver photon detector's deadtime, which we denote by $\tau_{hd}$. Photons that fall within the detector's deadtime are not detected. The number of detected photons in unit time that fall within the deadtime of heralding detector 1 or 2 can be expressed as

$$n_{di} = \frac{n_{ti}}{1 + n_{ti}\tau_{hr}}, \tag{12}$$

where

$$n_{ti} = \left[1 - e^{-(N_{dj} - N_{\text{XOR}})\tau_{hd}\beta_i\mu}\right]F. \tag{13}$$

Here, $i, j = 1, 2$ are labels for the heralding detectors, and $i \neq j$. The proportion of heralding photons that fall within the deadtime of a preceding photon can be expressed as

$$n_{ri} \equiv \frac{\Delta N_{di}}{N_{di}} n_{di}, \tag{14}$$

where

$$\Delta N_{di} = \frac{N_{ti}}{1 + N_{ti}\tau_{hr}} - \frac{N_{ti}}{1 + N_{ti}\tau_{hd}}. \tag{15}$$

The total number of heralded photons detected in unit time can thus be approximated by

$$N_{\text{detected}} = \alpha_s \left[ N_{\text{trigger}} - \alpha_s \sum_{i=1}^{2} (\Delta N_{di} + n_{di} - \alpha_s n_{ri}) \right]. \tag{16}$$

The above expression can be interpreted in a simple way. The first term is heralding efficiency multiplied by heralding rate, while the rest are correction terms taking into account the number of photons that would not be detected at the receiver if a preceding photon has been detected.

**Acknowledgment**

The authors thank J. Y. Yap for programming the CPLD chip.